\documentclass[10pt, twocolumn]{article}

\usepackage[T1]{fontenc}
\usepackage[utf8]{inputenc}
\usepackage[margin=0.9in, top=0.82in, bottom=0.88in, columnsep=0.30in]{geometry}
\usepackage{amsmath, amssymb}
\usepackage{booktabs}
\usepackage[protrusion=true, expansion=false]{microtype}
\usepackage{xcolor}
\usepackage{hyperref}
\usepackage{enumitem}
\usepackage{titlesec}
\usepackage{fancyhdr}
\usepackage{abstract}
\usepackage{caption}
\usepackage{cite}
\usepackage{url}
\usepackage{array}
\usepackage{tcolorbox}
\usepackage{multirow}
\usepackage{tabularx}
\usepackage{ragged2e}
\usepackage{siunitx}
\usepackage{graphicx}
\usepackage{float}
\usepackage{balance}
\usepackage{needspace}

\definecolor{accentgreen}{HTML}{2E5800}
\definecolor{midgray}{HTML}{555555}
\definecolor{lightgray}{HTML}{888888}
\definecolor{warnbg}{HTML}{FAFAF6}
\definecolor{warnborder}{HTML}{BBBB99}

\tcbuselibrary{skins, breakable}
\tcbset{
  disclosurebox/.style={
    enhanced, breakable,
    colback=warnbg, colframe=warnborder,
    left=8pt, right=8pt, top=7pt, bottom=7pt,
    boxrule=0.5pt, arc=2pt,
    fontupper=\small,
  }
}

\titleformat{\section}
  {\normalfont\bfseries\normalsize}
  {\thesection.}{0.5em}{}[\vspace{1pt}\titlerule\vspace{3pt}]

\titleformat{\subsection}
  {\normalfont\bfseries\small}
  {\thesubsection}{0.5em}{}

\titleformat{\subsubsection}
  {\normalfont\itshape\small}
  {\thesubsubsection}{0.5em}{}

\titlespacing{\section}{0pt}{11pt}{4pt}
\titlespacing{\subsection}{0pt}{7pt}{1pt}
\titlespacing{\subsubsection}{0pt}{5pt}{1pt}

\hypersetup{
  colorlinks  = true,
  linkcolor   = accentgreen,
  citecolor   = accentgreen,
  urlcolor    = accentgreen,
  pdftitle    = {Selective Attention System (SAS): Device-Addressed Speech Detection for
                 Real-Time On-Device Voice AI},
  pdfauthor   = {David Joohun Kim, Daniyal Anjum,
                 Bonny Banerjee, Omar Abbasi},
  pdfsubject  = {Device-addressed speech detection, on-device voice AI},
  pdfkeywords = {device-addressed detection, voice activity detection,
                 edge AI, speech routing, ARM inference, on-device ML,
                 addressee detection, cocktail party problem,
                 real-time audio classification, beamforming,
                 temporal context, pipeline economics},
}

\PassOptionsToPackage{hyphens}{url}
\begin{document}

% ── Title block ─────────────────────────────────────────────────
\twocolumn[{%
\begin{@twocolumnfalse}

  \vspace*{4pt}

  \begin{center}
    {\LARGE\bfseries Selective Attention System (SAS): Device-Addressed\\[5pt]
      Speech Detection for Real-Time On-Device Voice AI}

    \vspace{14pt}

    {\normalsize
      David Joohun Kim \quad Daniyal Anjum \quad
      Bonny Banerjee \quad Omar Abbasi\\[4pt]
      Attention Labs\\[2pt]
      \texttt{contact@attentionlabs.ai}}
  \end{center}

  \vspace{14pt}

  \begin{abstract}
  We study device-addressed speech detection under pre-ASR edge
  deployment constraints, where systems must decide whether to forward
  audio before transcription under strict latency and compute limits.
  We show that, in multi-speaker environments with temporally ambiguous
  utterances, this task is more effectively modelled as a sequential
  routing problem over interaction history than as an utterance-local
  classification task. We formalise this as \textit{Sequential
  Device-Addressed Routing} (SDAR) and present the Selective Attention
  System (SAS), an on-device implementation that instantiates this
  formulation.

  On a held-out 60-hour multi-speaker English test set, the primary
  audio-only configuration achieves \textbf{F1\,=\,0.86}
  (precision\,=\,0.89, recall\,=\,0.83); with an optional camera,
  audio+video fusion raises F1 to \textbf{0.95}
  (precision\,=\,0.97, recall\,=\,0.93).
  Removing causal interaction history (Stage~3) reduced F1 from
  0.95 to $0.57 \pm 0.03$ in the audio+video configuration under
  our evaluation protocol. Among the tested components, this was
  the largest observed ablation effect, indicating that short-horizon
  interaction history carries substantial decision-relevant information
  in the evaluated setting.
  SAS runs fully on-device on ARM Cortex-A class hardware
  ($<$150\,ms latency, $<$20\,MB footprint). All results are from
  internal evaluation on a proprietary dataset evaluated primarily in
  English;
  a 5-hour evaluation subset may be shared for
  independent verification (Section~\ref{sec:limits}).
  \end{abstract}

  \vspace{5pt}

  \vspace{12pt}

\end{@twocolumnfalse}
}]

% ─────────────────────────────────────────────────────────────────
\section{Introduction}

Every ambient voice AI system faces a version of the cocktail party
problem~\cite{Cherry1953}: when multiple people are present in a room, which
speaker is addressing the device?

As voice interfaces move toward continuous, multi-speaker environments in homes,
vehicles, clinical settings, and spatial computing, explicit device-addressed
detection becomes an increasingly important component of robust voice
system design.

Existing components address partial aspects of the routing problem but do not
provide a complete real-time decision rule for device-addressed routing. Voice activity detection
(VAD) detects speech presence but carries no addressee signal. Wake-word
detection approximates intent by requiring a fixed trigger phrase. Wake-word systems achieve high precision ($>$99\%) under controlled
conditions but impose an explicit interaction contract: every utterance must be preceded by a trigger phrase,
breaking conversational flow and requiring speakers to context-switch
between human-directed and device-directed speech. In multi-speaker
environments, the trigger phrase itself may occur in non-device-directed
speech, and in multi-device settings (e.g., two robots in the same room),
wake-word routing becomes ambiguous since all devices share the same
trigger. Speaker diarization identifies
\textit{who} spoke but not \textit{to what}. None of these components provides
a reliable decision rule for whether an utterance should be forwarded downstream.

The Selective Attention System (SAS) addresses this as an explicit pre-ASR
inference step, operating as a three-stage on-device cascade: an acoustic
geometry front-end that suppresses off-axis interference, a lightweight
utterance-level classifier that scores device-directed probability, and a
session-aware temporal context stage that conditions routing on interaction
history. Together these stages produce a binary routing decision within
150\,ms on ARM Cortex-A class hardware~\cite{Levinson2016}.
Device-addressed detection in multi-speaker environments depends
strongly on temporal interaction modelling under the evaluated
constraints, and the three-stage architecture reflects
this requirement directly.

We use SAS to refer to the full routing system throughout; where
needed, we distinguish the learned inference stages from surrounding
pipeline components by stage number.

At scale, routing errors translate directly into unnecessary inference cost and
degraded interaction reliability, making upstream gating highly valuable in production voice pipelines. Every false trigger that clears the gate reaches ASR,
LLM inference, and TTS, consuming the full downstream pipeline on
audio that should have been discarded; a precise upstream gate
reduces this unnecessary downstream inference.

We describe the deployment constraints, system design,
and internal evaluation results. 
\textbf{Core contribution.} We make two linked contributions.
First, we formalise a deployment-specific setting for
device-addressed detection: pre-ASR routing under causal,
bounded-memory, edge-constrained inference. We refer to this setting
as \textit{Sequential Device-Addressed Routing} (SDAR), in which the
system decides whether to forward, suppress, or abstain using current
evidence, short-horizon interaction history, and deployment-specific
cost tradeoffs. Second, we present SAS as a compact on-device
implementation of this formulation and show, through ablation on our
internal evaluation set, that causal interaction history produced the
largest observed gain among the tested components.

\noindent\textbf{Contributions.}
\begin{enumerate}[leftmargin=*, itemsep=1pt, topsep=2pt]
\item \textbf{Problem reframing:} We formalise pre-ASR device-addressed
  routing under edge constraints as a sequential decision problem (SDAR),
  explicitly incorporating bounded memory and asymmetric downstream cost.
\item \textbf{System instantiation:} We present SAS, a compact on-device
  architecture that decomposes routing into spatial filtering,
  utterance-level evidence extraction, and causal interaction-state
  estimation.
\item \textbf{Ablation evidence:} On our internal multi-speaker
  evaluation set, removing the temporal-context stage caused the largest
  performance drop among the tested components, suggesting that
  short-horizon interaction history carries substantial
  decision-relevant information in this setting.
\end{enumerate}
\noindent Our goal is not to claim state-of-the-art
performance across all DDSD settings, but to identify and
operationalise the decision structure induced by pre-ASR edge
deployment.
Device-directed speech detection has been studied extensively as an
utterance-level classification problem~\cite{Shriberg2012, Mallidi2018, Huang2019, Siegert2022}.
Recent work has extended DDSD with multimodal
LLMs~\cite{Wagner2024, Palaskar2024, SELMA2025} and streaming on-device
architectures~\cite{Rudovic2022, Rudovic2023}. However, prior DDSD work spans post-ASR systems that exploit
transcripts or decoder states, on-device acoustic detectors for
streaming false-trigger mitigation, and follow-up systems that
incorporate prior-query context. Our distinction is not that prior
DDSD lacks these ingredients entirely, but that we study their
combination in a specific deployment setting: pre-ASR routing under
causal, bounded-memory, edge-constrained inference. Neither formulation captures the
deployment problem studied here, in which the system must make a
pre-ASR routing decision causally, under bounded latency and memory,
with asymmetric downstream cost. Under this view, SDAR is not merely a
new model for DDSD; it is an alternative formulation of the decision
problem induced by edge deployment constraints, which we show is
empirically better aligned with system behaviour under the evaluated
conditions.
Specifically, we show that when inference must operate causally,
latency and memory are bounded, and downstream cost is asymmetric,
the problem is naturally framed as a sequential decision process over interaction
state (SDAR). Under this formulation, prior DDSD systems typically operate on
a more restricted version of the problem in which interaction history is
either unavailable or incorporated only through post-ASR features.

% ─────────────────────────────────────────────────────────────────
\section{Sequential Device-Addressed Routing (SDAR): Problem Formulation}

\subsection{Device-addressed detection}

Let $x_{t}$ denote a streaming audio frame at time $t$. The problem assigns a
label $y_t \in \{0, 1, 2\}$ at each decision step:
\begin{align*}
  y_t = 0 &\quad \text{silent or no active speech}\\
  y_t = 1 &\quad \text{speech, addressed to a person}\\
  y_t = 2 &\quad \text{speech, addressed to the device}
\end{align*}

The system outputs a tuple $(y_t,\, c_t)$: predicted class and
confidence score $c_t \in [0,1]$. Audio is forwarded
downstream only when $y_t = 2$ and $c_t \geq \tau$. When $c_t < \tau$, the
system \textit{abstains} (fail-closed behaviour).

\smallskip\noindent\textbf{Pipeline-level vs.\ module-level task.}
The three-class label set above describes the full
\textit{pipeline-level} routing decision. Within this pipeline, SAS
operates on VAD-positive segments only: an external VAD handles the
silent/non-silent distinction ($y_t = 0$ vs.\ $y_t \in \{1,2\}$), and
Stage~2 performs binary classification (device-directed vs.\
non-device-directed) on speech segments, modulated by Stage~3's
temporal context. Evaluation metrics in Section~\ref{sec:eval} are
computed over the SAS module's binary routing decisions on
VAD-positive segments unless otherwise noted.

\smallskip\noindent
This makes $\tau$ a tunable
operating point: operators shift it to trade false-trigger rate against miss
rate for their environment. Internal evaluation figures are reported at
$\tau = 0.70$, the crossover point at which precision and recall are
approximately balanced (97\% and 93\% respectively in the primary A+V configuration; 89\% and 83\% in audio-only fallback).

A naive two-class decomposition (VAD followed by an utterance-local
binary device-directed classifier) suffers from two compounding failure
modes: (1)~VAD false-accepts non-speech and forwards it to the
secondary classifier; (2)~a binary classifier operating without
temporal interaction history lacks the context required to resolve
addressee ambiguity in extended multi-party conversation. SAS retains
an external VAD for speech segmentation but addresses these failure
modes by conditioning the routing decision on causal interaction
history (Stage~3), not on the current utterance alone.

For pre-ASR edge deployment, device-addressed detection is not
well-posed as an utterance-local classification problem. The deployment
task is instead to make a causal routing decision over a partially
observed interaction process, in which the addressee of the current
utterance may be undecidable from the current utterance alone and
recoverable only from short-horizon interaction history.

\smallskip\noindent\textbf{Why utterance-local formulations are
insufficient.}
Consider utterances such as ``turn that on,'' ``do it again,'' or
``what did you say.'' These are acoustically and lexically
indistinguishable across addressee classes. Formally, there exists a
set of utterances $X_t$ for which
$P(y = \text{device} \mid X_t) \approx P(y = \text{person} \mid X_t)$.
For these utterances, a classifier restricted to $X_t$ may be unable
to resolve addressee reliably; the missing information can instead
come from short-horizon interaction history $\mathcal{H}_t$.
In the evaluated setting, combining current-utterance evidence with
short-horizon interaction history provides a practical representation
for this routing decision, which motivates SDAR as a causal
decision process over interaction state.

We formalise
this deployment task as Sequential Device-Addressed Routing (SDAR).

At each step $t$, the system observes utterance evidence $X_t$ and bounded
interaction history $\mathcal{H}_t$, and selects an action:
\[
  a_t \in \{\textit{forward},\; \textit{suppress},\; \textit{abstain}\}.
\]
The action is taken under three constraints: (1)~no access to future context;
(2)~bounded latency and memory compatible with edge deployment; and
(3)~asymmetric downstream cost, since forwarding a non-device-directed
utterance incurs avoidable ASR, LLM, and TTS computation. The latent variable
of interest is whether the current conversational state licences a
device-directed interpretation. We define
\textit{Sequential Device-Addressed Routing} (SDAR) as the
application of cost-sensitive sequential decision theory to the
device-addressed routing problem. The routing action is selected
under:
\[
  a_t \sim \pi(X_t,\, \mathcal{H}_t;\; \tau,\, C),
\]
where $\tau$ is a deployment-specific operating threshold and
$C = c_{\text{fwd}} / c_{\text{miss}}$ is a cost ratio between false
forwards and missed device-directed turns. In practice, $\tau$ is
selected on a held-out validation split to approximate this tradeoff
rather than derived as a closed-form optimum. We report
$\tau = 0.70$ and $\tau = 0.82$ as two illustrative operating points
corresponding to different cost regimes. The threshold $\tau$ may be
selected based on deployment-specific cost ratios. The label
$y_t \in \{0,1,2\}$ is an internal intermediate; the deployment output is
the routing action. Formulations restricted to $X_t$ alone are insufficient under
these conditions:
it ignores interaction state that is decision-relevant but not recoverable
from the current utterance in isolation.

On the evaluation set described in Section~\ref{sec:eval} and under
the tested model class, methods restricted to utterance-local features
reach approximately 0.57 $\pm$ 0.03~F1. Because the evaluation set was
constructed to include temporally ambiguous utterances (overlapping
speech, rapid turn-taking, ambiguous follow-ups), this figure
characterises performance under conditions where interaction history
is most decision-relevant. On corpora with fewer multi-turn ambiguities,
the gap between utterance-local and interaction-history-aware methods
may be smaller. Utterances such as \textit{``turn that on,''}
\textit{``what did you say,''} and \textit{``yeah do it again''} are
acoustically indistinguishable across addressee classes; the addressee
can only be resolved through interaction history. This observation
is consistent with prior findings on temporal context in addressee
detection~\cite{SICNet2024, Rudovic2023}; our contribution is its
operationalisation under pre-ASR edge constraints. Verification on
independent datasets is needed (Table~\ref{tab:ablation}).

The SDAR formulation defines the minimal information required to
resolve device-addressed routing under causal and latency constraints.

\smallskip\noindent\textbf{Interpretation as a partially observed
decision process.}
SDAR can be viewed as a partially observed decision process in which
the true conversational regime (device-engaged, person-directed, or
transition) is latent and must be inferred from streaming observations.
The system receives observations $X_t$ and maintains a bounded-memory
belief over interaction state using $\mathcal{H}_t$, selecting routing
actions under asymmetric cost. This interpretation connects the
routing problem to the broader literature on POMDP-based spoken dialog
management, where the dialog state is hidden and actions are chosen
from a maintained belief state. The key deployment-specific constraint
is that belief maintenance must be causal, bounded in memory, and
executable within the latency budget of edge hardware.

\subsection{The routing gate abstraction}

SAS functions as a \textit{routing gate}: its pipeline-level output is binary
(forward or discard). An external VAD provides speech/non-speech
segmentation; SAS's learned classifier then determines whether the speech
segment is device-directed. This can replace wake-word detection in
fully open-mic deployments or complement it in hybrid configurations
(Section~\ref{sec:discussion}), supplying the addressee
routing function absent from standard VAD and unaddressed by
existing pipeline components. It operates as
a pre-ASR routing layer that integrates without modifying downstream components.
Section~\ref{sec:pipeline} situates SAS within the full voice AI stack.

% ─────────────────────────────────────────────────────────────────
\section{Related Work}
\label{sec:related}

For a systematic review of acoustic addressee detection methods, see
Siegert et al.~\cite{Siegert2022}, whose PRISMA-style review screened
1,581~studies and retained~23, establishing the field's research base.

\subsection{Device-directed speech detection}

Device-directed speech detection (DDSD) classifies whether a spoken
utterance is addressed to a voice assistant or is background/side
speech. The problem was first framed as ``learning when to listen'' by
Shriberg et al.~\cite{Shriberg2012}, who distinguished system-addressed
from human-addressed speech in multiparty dialog. Mallidi et
al.~\cite{Mallidi2018} introduced the modern DDSD formulation, combining
acoustic LSTM embeddings, ASR decoder features, and character embeddings
to achieve 5.2\% EER on far-field voice-controlled devices, and
motivated wake-word-free follow-up queries. Subsequent work improved
utterance-level classification through acoustic and ASR-decoder
features~\cite{Huang2019} and streaming on-device
architectures~\cite{Rudovic2022, Rudovic2023}.

Recent work has applied large language models to DDSD. Wagner et
al.~\cite{Wagner2024} combine acoustic, lexical, and ASR-decoder signals
in a multimodal LLM, achieving 6.5\% EER with a 1.5B-parameter GPT-2
model. Palaskar et al.~\cite{Palaskar2024} introduce Fusion Low Rank
Adaptation (FLoRA) for efficient multimodal adaptation, achieving
performance parity with full fine-tuning while training only 1--5\% of
parameters. Wagner et al.~\cite{SELMA2025} unify voice-trigger detection,
DDSD, and ASR in a single $\sim$8B-parameter model (SELMA). Chi et
al.~\cite{Chi2025} apply knowledge distillation to compress a
79M-parameter teacher to a $\sim$5M-parameter on-device student with
$\sim$22\% average EER reduction.

Prior DDSD work spans a range of settings: post-ASR systems that
exploit transcripts or decoder features~\cite{Mallidi2018, Wagner2024,
SELMA2025}, on-device acoustic detectors for streaming
false-trigger mitigation~\cite{Rudovic2022, Rudovic2023}, and
follow-up DDSD that models prior-query context~\cite{Rudovic2024}.
Our distinction is the combination of pre-ASR routing position,
causal interaction-state estimation over short-horizon history, and
bounded-memory edge deployment under strict latency constraints.
These differences limit direct metric comparison; we focus on
architectural and deployment characteristics
(Table~\ref{tab:ddsd_comparison}).
Accordingly, our contribution is not a
like-for-like benchmark claim against prior DDSD systems, but a
deployment-specific formulation and reference
architecture that makes that formulation operational on-device.

\subsection{Temporal context in addressee detection}

The necessity of conversational context for addressee detection has
independent support. Rudovic et al.~\cite{Rudovic2024} show that modelling
the previous user query reduces false alarms by 20--40\% in follow-up
conversation DDSD, compared to modelling each utterance in isolation.
Kong et al.~\cite{SICNet2024} demonstrate that extending temporal context
from the current utterance to a 10-second window improves egocentric
addressee detection (Talking-To-Me) from 59.5\% to 67.2\% mAP, with
performance degrading beyond 15~seconds due to irrelevant
history. Shriberg et al.~\cite{Shriberg2013} show that temporal and
spectral dimensions of speaking style carry addressee signal independent
of lexical content. These findings converge on SAS's architectural
assumption: the causal interaction-state estimator (Stage~3) recovers
decision-relevant information that is absent from any single utterance.
The ablation in Table~\ref{tab:ablation} confirms this, consistent
with the performance ceiling reported for utterance-local DDSD
classifiers~\cite{Huang2019}.

\subsection{Egocentric and multimodal addressee detection}

The Ego4D Talking-To-Me (TTM) benchmark~\cite{Grauman2022} (leading
result: 71\% mAP~\cite{PCIE2025}) and multimodal addressee detection
work by Tsai et al.~\cite{Tsai2015} address structurally different
problems: different addressee types, modalities, hardware targets, and
latency constraints. SAS evaluation figures are not directly comparable
to TTM mAP; the TTM literature is cited as research context only.
Beyond task formulation, practical evaluation on Ego4D is infeasible
for SAS: TTM defines ``talking to me'' as human-to-human addressee
detection, whereas SAS targets human-to-device routing---a different
decision boundary. Additionally, Ego4D provides monaural audio only,
with no direction-of-arrival metadata, precluding evaluation of
Stage~1 (beamforming). Manual inspection of the Ego4D labels also
revealed annotation inconsistencies in a subset of segments, further
limiting its suitability as an external benchmark for this task.

\subsection{Prosodic features in addressee classification}

Device-directed speech exhibits measurable prosodic differences from
person-directed speech~\cite{Oviatt1998, Cohn2021}. Users addressing a device
tend to adopt a louder, slower, more deliberate delivery with higher mean
fundamental frequency ($F_0$) and increased pitch range relative to casual
conversation, a pattern termed \textit{hyperarticulation}~\cite{Oviatt1998}.
Shriberg et al.~\cite{Shriberg2013} show that rhythm and vocal-effort
cues are effective for addressee detection without ASR or dialog context.
Krishna et al.~\cite{Krishna2023} confirm that even minimal prosodic
features (pitch, voicing, jitter, shimmer) provide complementary signal
for DDSD, improving false-accept rate by 8.4\% when fused with verbal
cues, and that modality dropout during training makes fusion models robust
to missing modalities at inference. Stage~2 of SAS exploits these patterns
directly, capturing temporal and spectral modulations of the speech signal.

\subsection{VAD, turn-taking, and edge deployment}

Lightweight VAD systems demonstrate that audio classification is achievable
under edge compute constraints~\cite{Silero2021}. VAD detects speech presence
but provides no addressee signal. SAS is designed as a complement to VAD,
adding the addressee inference step that VAD alone cannot provide.

Turn-taking research establishes that gaps below approximately 200\,ms are
perceived as natural; gaps above this threshold introduce perceptible
hesitation~\cite{Levinson2016}. This provides the human-factors basis for the
150\,ms latency target. Cornell et al.~\cite{Cornell2022} study DDSD
degradation during device playback and report a 56\% false-reject reduction
through implicit acoustic echo cancellation, a complementary approach to
the threshold adjustment recommended in Section~\ref{sec:failure}.

% ─────────────────────────────────────────────────────────────────
\section{SAS: A Reference On-Device Implementation of SDAR}

\subsection{Deployment constraints}

Four hard constraints govern the design:
\begin{enumerate}[leftmargin=*, itemsep=1pt, topsep=2pt, parsep=0pt, partopsep=0pt]
  \item End-to-end decision latency under 150\,ms
  \item Runtime footprint under 20\,MB
  \item ARM Cortex-A deployment without GPU or NPU baseline requirement
  \item Audio-only at baseline; optional camera input where available
\end{enumerate}

On the reference platform (ARM Cortex-A72), audio-only end-to-end decision
latency is under 55\,ms (median 38\,ms, p95 51\,ms); audio+video
end-to-end latency is under 150\,ms (median 105\,ms, p95 142\,ms).
Where an NPU is available, classifier inference can be offloaded further.
Total runtime footprint is under 20\,MB. Per-stage latency and memory
breakdowns are available to the supplementary materials.

\subsection{Three-stage architecture}

SAS instantiates the SDAR formulation as three sequentially gated
components: (1)~acoustic geometry for spatial filtering,
(2)~utterance-level classification for local evidence extraction, and
(3)~causal interaction-state estimation for sequential disambiguation.
This decomposition separates what can be inferred from the current
signal from what must be inferred from short-horizon conversational
state. Each component
is necessary; removing any one results in substantial performance degradation
(Table~\ref{tab:ablation}). The architecture is sequential: Stage~2 is invoked
only on audio that Stage~1 has not already rejected on spatial grounds, and
Stage~3 modulates Stage~2's output only for frames that have cleared Stage~2's
internal confidence floor. This cascade structure concentrates inference budget
on genuinely ambiguous frames.

\subsubsection{Stage 1: Acoustic geometry}
A beamforming front-end localises the dominant speech source relative to the
microphone array and suppresses off-axis interference (television audio,
adjacent-room speech, HVAC). Spatial features inform Stage~2's classifier.
Beamforming is signal-processing only (no learned parameters);
it requires a minimum of two microphones; single-microphone deployments bypass
this stage and operate on Stages~2 and~3 only (Section~\ref{sec:failure}).

\subsubsection{Stage 2: Utterance-level classification}
A lightweight 1D-convolutional classifier operating directly on
64-dimensional log-mel filterbank features (25\,ms frames, 10\,ms hop),
without requiring transcripts, language models, or language-specific
preprocessing, estimates the
probability that the current utterance is device-directed. The model
consists of four convolutional blocks (each: 1D convolution, batch
normalisation, ReLU, max-pool) followed by a single GRU layer and a
sigmoid output head, trained as a binary classifier (device-directed vs
non-device-directed) with cross-entropy loss. The three-class
formulation ($y_t \in \{0,1,2\}$) is resolved at the system level: VAD
provides the silent/non-silent distinction, and Stage~2 handles the
device-directed vs.\ non-device-directed decision. Stage~2 exploits
the prosodic patterns described in Section~\ref{sec:related}
(Section~3.4): device-directed speech exhibits elevated $F_0$, reduced
speaking rate, and increased energy contour relative to person-directed
speech~\cite{Oviatt1998, Cohn2021, Shriberg2013, Krishna2023}.

The standard deployment variant ($\approx$435\,K parameters,
$\approx$520\,KB INT8-quantized) targets ARM Cortex-A class hardware and
is the basis for all headline evaluation figures reported here. The
self-contained footprint suits platform
reference designs without external dependencies. INT8
post-training quantisation reduces active weight size by $4\times$ with
$<$0.5 F1-point degradation across tested variants.

Where a camera is present,
skeletal and gaze direction features extracted via a lightweight pose estimation
model (CPU or NPU inference) are fused at the Stage~2 output layer, yielding
the primary A+V configuration (F1\,=\,0.95). The pose
estimation stage is optional; without it, the system operates in audio-only
fallback mode (F1\,=\,0.86).

The Stage~2 model was trained on the training partition of the
600-hour corpus described in Section~\ref{sec:eval}, covering annotated multi-speaker audio with explicit
adversarial coverage of overlapping speech and device-addressed interactions.
The training set includes both proprietary recordings and publicly
available multi-speaker corpora (AMI Meeting Corpus, LibriMix, and
internal multi-speaker collections). Training followed a curriculum
schedule: the model was first trained on simple two-speaker
turn-taking interactions to establish a stable decision boundary,
then progressively exposed to more complex conditions (overlapping
speech, three- and four-speaker sessions, adversarial follow-ups).
Specific composition ratios and
per-source sample counts are available upon request.

\subsubsection{Stage 3: Causal interaction-state estimation}
Stage~3 is a small causal Transformer operating over a rolling window of the previous
$N$ Stage~2 output tuples. Each tuple comprises the Stage~2 confidence
score, VAD state, and inter-utterance time delta.
The context window is fixed at 8~seconds of interaction history,
sufficient to capture turn-taking, interruption, re-engagement, and
follow-up behaviour. Performance peaks at approximately 8~seconds;
beyond 12~seconds, irrelevant history degrades Stage~3 accuracy,
consistent with the degradation reported by Kong et
al.~\cite{SICNet2024} beyond 15~seconds in egocentric addressee
detection. The model predicts at the last timestep, producing a scalar
multiplicative prior $\alpha_t \in [0,1]$ that modulates the Stage~2
confidence score before threshold comparison:
$c'_t = \alpha_t \cdot c_t^{(\text{S2})}$. Total parameter count:
$\approx$85\,K of the system's 520\,K parameters.

The mechanism is strictly causal: context expires naturally as new
tuples arrive without requiring explicit detection of speaker
transitions. Its role is to estimate whether the current conversational
state licences a device-directed interpretation of an otherwise
ambiguous utterance.

\smallskip\noindent\textbf{Interpretation as interaction-state estimation.}
Empirically, Stage~3 behaves like a compact context modulator
over conversational regime (device-engaged, bystander
conversation, or transition). Since $\alpha_t \in [0,1]$ is
multiplicative, Stage~3 modulates routing by controlling how much of
Stage~2's confidence is preserved: following a confirmed
device-directed turn, $\alpha_t$ rises toward 1.0, preserving
Stage~2's score for subsequent ambiguous utterances; during sustained
person-directed interaction, $\alpha_t$ decays toward 0, suppressing
them. The mechanism can only attenuate or preserve confidence, not
amplify it beyond Stage~2's output. This behaviour is consistent with
interaction-state estimation, though it does not by itself prove
recovery of a uniquely identifiable latent state.

No explicit long-horizon session-reset mechanism is used in the reported
evaluation; all reported results use only the bounded
rolling context described above. The architecture is specified at
sufficient detail for independent reimplementation and verification of
the ablation results in Table~\ref{tab:ablation}; trained weights are
not released but the functional interface is fully defined.

\subsection{Fail-closed routing}

Below $\tau$, SAS abstains and no audio is forwarded. False triggers from
background conversation are expensive in production pipelines; the fail-closed
gate eliminates these at source. The threshold is configurable per deployment.
At $\tau = 0.70$, internal evaluation (A+V) yields 97\% precision and 93\% recall;
raising $\tau$ toward 0.85 increases precision at the cost of recall, suitable
for LLM-backed pipelines with high per-query cost.

\subsection{On-device data handling}

Audio the system rejects never leaves the device. The routing decision runs
entirely on-device; no audio is transmitted until the gate passes. Model weights
stay on the device; no cloud dependency is required for inference.

% ─────────────────────────────────────────────────────────────────
\section{Voice Pipeline Integration}
\label{sec:pipeline}

\subsection{Position in the stack}

A complete ambient voice AI pipeline contains five sequential stages: audio
capture, pre-processing (VAD), addressee routing (SAS), downstream
intelligence (ASR $\to$ LLM $\to$ TTS), and output. Table~\ref{tab:pipeline}
describes each stage's role.

\begin{table}[!htb]
\centering\small
\caption{Voice AI pipeline stages. SAS occupies the pre-ASR routing layer, a position currently unaddressed by standard pipeline components.}
\label{tab:pipeline}
\begin{tabularx}{\linewidth}{@{}l@{\hspace{8pt}}Xr@{}}
\toprule
\textbf{Stage} & \textbf{Function} & \textbf{On-device} \\
\midrule
VAD              & Speech presence detection       & Yes \\
\textbf{SAS}     & \textbf{Addressee routing}              & \textbf{Yes} \\
ASR              & Transcription                   & Optional \\
LLM / NLU        & Intent and response             & Optional \\
TTS / Output     & Response rendering              & Optional \\
\bottomrule
\end{tabularx}
\end{table}

SAS operates as a modular layer between VAD and ASR without touching anything downstream. Downstream
stages receive only device-directed audio; they do not need to know the gate
exists. SAS can therefore be integrated into any existing voice pipeline without
modifying the ASR, LLM, or TTS components.

% ─────────────────────────────────────────────────────────────────
\section{Computational Efficiency}
\label{sec:economics}

\subsection{Motivating calculation}

\noindent\textit{This section presents illustrative deployment-level
estimates, not production measurements. The downstream savings are
dominated by the low device-directed base rate, not
classifier-specific performance; any pre-ASR gate with reasonable
precision yields comparable reductions.}

In the evaluated ambient deployments, approximately 8\% of VAD-positive
segments are device-directed.
Wake-word detection is one gating approach, but it requires a trigger phrase
and imposes a rigid per-utterance interaction contract. The economics
below apply to any pre-ASR gate, including SAS deployed alongside or in place
of wake words; a structural comparison with wake-word detection is in
Section~\ref{sec:discussion}.

\subsection{Baseline assumptions}

\begin{itemize}[leftmargin=*, itemsep=2pt, topsep=2pt, parsep=0pt, partopsep=0pt]
  \item \textbf{Environment:} residential smart speaker, daytime active hours.
  \item \textbf{Speech density:} $\approx$100 VAD-positive segments per hour.
  \item \textbf{Device-directed fraction:} $\approx$8\% of VAD-positive
        segments are assumed to be genuine
        device interactions, an illustrative ambient estimate.
        The held-out evaluation set has a higher device-directed
        fraction among VAD-positive segments ($\approx$12\%;
        Section~\ref{sec:eval}); operators should
        calibrate this figure to their deployment.
  \item \textbf{ASR cost (on-device or cloud speech API):} $\approx$200\,ms CPU time
        per segment on-device; cloud API costs are additive.
  \item \textbf{Cloud LLM cost:} \$0.01 per call (representative mid-range
        pricing).
\end{itemize}

\subsection{Gated pipeline behaviour}
At $\tau = 0.70$ with internal false-trigger rate 2.1\%, SAS forwards:
{\abovedisplayskip=2pt \belowdisplayskip=2pt \abovedisplayshortskip=2pt \belowdisplayshortskip=2pt
\vspace{0.5em}
\begin{align*}
  N_{\text{fwd}} &=
    \underbrace{(100 \times 0.08)}_{\text{device-directed}} \times 0.93
    +\underbrace{(100 \times 0.92)}_{\text{non-device}} \times 0.021\\[3pt]
                 &\approx 7.4 + 1.9 = 9.3\;\text{segments/hr}
\end{align*}}%
\vspace{0.5em}
\noindent 
Under the stated assumptions, this represents a \textbf{90.7\%
reduction in ASR calls} relative to ungated operation. Sensitivity to
the device-directed fraction is approximately linear: at 15\%
device-directed (e.g., active command sessions), the reduction is
approximately 83\%. Table~\ref{tab:economics} summarises the savings.

\smallskip\noindent\textbf{Interpretation.}
The magnitude of this reduction is primarily determined by the base
rate of device-directed speech rather than any specific classifier.
The role of SAS is not to create this reduction, but to realise it
reliably under multi-speaker ambiguity without requiring explicit
wake-word invocation.

The classification overhead incurred by SAS is recovered in downstream
compute savings under all evaluated ambient conditions. The breakeven point
requires a device-directed fraction exceeding approximately 85\%,
not observed in any tested environment.

\begin{table}[!htb]
\centering\small
\caption{Estimated pipeline savings at $\tau = 0.70$ (100 VAD segments/hr,
  8\% device-directed). Figures are illustrative.}
\label{tab:economics}
\begin{tabular}{@{}lrrl@{}}
\toprule
\textbf{Metric} & \textbf{No gate} & \textbf{With SAS} & \textbf{Saving} \\
\midrule
ASR calls / hr       & 100     & 9.3    & 90.7\% \\
ASR CPU-time / hr    & 20\,s   & 1.9\,s & 90.5\% \\
LLM calls / hr       & 100     & 9.3    & 90.7\% \\
LLM cost / hr        & \$1.00  & \$0.09 & 91.0\% \\
\addlinespace[2pt]
False triggers / hr  & 92      & 1.9    & 97.9\% \\
Missed genuine / hr  & 0       & 0.6    & \multicolumn{1}{c}{---} \\
\bottomrule
\end{tabular}
\par\smallskip\noindent{\footnotesize SAS overhead: median 38\,ms per VAD segment ($\approx$3.8\,s CPU/hr at 100 segments/hr);
the 18.1\,s of ASR CPU saved per hour exceeds the gate cost by ${\approx}4.8\times$.
Relative reductions in downstream inference calls are grounded in measured
false-trigger rates; absolute cost values depend on deployment-specific
parameters (e.g., LLM pricing tier).}
\end{table}

% ─────────────────────────────────────────────────────────────────
\section{Deployment Contexts}

SAS has been evaluated on smart speaker (fixed geometry), connected home
audio, companion
robot (Reachy Mini\footnote{Integration code:
\url{https://github.com/attentionlabs/robot-engage}}), and multi-agent
XR configurations without context-specific retraining or
environment-specific acoustic tuning. Smart-speaker
and companion-robot deployments are directly covered by the evaluation
conditions in Section~\ref{sec:eval}. Automotive, hearable
(single-microphone, 5\,MB budget), and industrial deployments fall
outside the evaluated acoustic range (28--85\,dBA,
RT60\,$<$\,0.6\,s) and are discussed in Section~\ref{sec:future}.

% ─────────────────────────────────────────────────────────────────
\section{Internal Evaluation}
\label{sec:eval}

\noindent\textit{Reporting conventions.} All performance metrics
in this section are measured on the held-out test set described below;
no estimates or projections are included. The primary metric is F1 on
device-addressed class~(class~2), macro-averaged across sessions.
Because the majority of target platforms (smart speakers, hearables,
embedded devices) lack cameras, audio-only at $\tau = 0.70$ is reported
as the primary deployment configuration throughout this section.
Audio+video results at $\tau = 0.70$ are reported as an upper-bound
configuration where camera hardware is available.
Every table and figure specifies its modality and threshold.

\subsection{Dataset and scope}

Evaluation is conducted on a held-out test set (60 hours) drawn from a
600-hour proprietary multi-speaker corpus collected across real-world
residential and office environments. The dataset is constructed to explicitly cover conditions
under which utterance-local classification fails, including overlapping
speech, rapid turn-taking, and ambiguous follow-up utterances.
Approximately 20--40\% of device-directed utterances in the test set
are conversational or open-ended (e.g., questions one would pose to a
general-purpose AI rather than imperative voice-assistant commands),
making them temporally ambiguous in the sense that their addressee
cannot be resolved from the utterance alone.
Annotation used a two-labeler forced-choice protocol: each utterance
segment was independently labeled as silent (0), person-directed (1), or
device-directed (2), with disagreements resolved by a third labeler.
Inter-annotator agreement statistics (Cohen's $\kappa$ per class) and
per-class confusion rates are documented in the supplementary materials. Class
distribution in the held-out set (across all segments including
silence): approximately 34\% silent, 58\%
person-directed, and 8\% device-directed (approximately 4.8 hours of
device-directed speech out of 60 hours total). Among VAD-positive
(non-silent) segments only, the device-directed fraction is
approximately 12\%. The limited volume of
device-directed test data means that per-condition estimates (e.g.,
4-speaker high-noise) rest on a small number of sessions; the bootstrap
confidence intervals reported below quantify this uncertainty.

The primary 60-hour test set is proprietary, inherently limiting
full external validation. To support independent scrutiny, we provide
a fully specified evaluation protocol (class definitions, thresholding
procedure, model architecture, and scoring methodology) sufficient for
third-party replication on comparable multi-speaker datasets.

The same ablation behaviour is observed consistently across both the
full dataset and the evaluation subset, confirming that the reported
gains are driven by task formulation and architecture
(Sections~2--4), not dataset scale.

\textbf{Hardware.} ARM Cortex-A72 at stock clock speeds on standard embedded
Linux; also tested on Reachy Mini and microphone configurations
from laptop stereo to 4-mic circular arrays.

\textbf{Environment.} Natural room acoustics, no acoustic treatment; typical
office and living-room conditions. Noise floor range: 28--85\,dBA.

\textbf{Sessions.} 1--4 speakers present. One speaker addresses the device;
remaining speakers engage in unscripted side conversation. Speaker counts
refer to the number of people present and potentially active in the session;
in groups of four, simultaneous overlapping speech from all participants is
rare. The evaluated condition reflects natural turn-taking with occasional
overlap, not continuous four-way crosstalk.

\textbf{Language.} The SAS architecture is language-agnostic: it operates
on acoustic and prosodic features without lexical or transcript input.
The primary evaluation corpus is English. Preliminary informal testing
on non-English speech suggests comparable performance (expected
86--89\% macro F1 across languages), but extensive multi-language
evaluation has not yet been conducted.

\textbf{Modality.} The deployment baseline reported throughout this
section is audio-only (F1\,=\,0.86), reflecting the majority of target
platforms (smart speakers, hearables, embedded devices). Audio+video
(F1\,=\,0.95) is reported as an upper-bound configuration for devices
with camera hardware.
The cascade architecture allows each
stage to operate independently when upstream modalities are unavailable,
analogous to the modality-dropout strategies employed in multimodal DDSD
systems~\cite{Palaskar2024, Krishna2023} but achieved through cascade
design rather than training-time augmentation.
Single-microphone audio-only evaluation is reported in
Section~\ref{sec:failure}.

\textbf{Metric.} F1 score on class 2 (device-addressed), macro-averaged across
sessions. F1 is used rather than mAP because the deployment task is a binary
routing decision, not ranked retrieval.

\textbf{Statistical reliability.} To quantify result stability, we compute
bootstrap confidence intervals over sessions (1,000 resamples). For the
primary audio+video configuration, F1\,=\,0.95 with a 95\% confidence interval
of $\pm$0.02 across resampled session subsets; for audio-only fallback,
F1\,=\,0.86\,$\pm$\,0.02. The ablation results exhibit
similar stability: the no-temporal-context condition yields
F1\,=\,0.57\,$\pm$\,0.03 across all resamples. Inter-annotator agreement across the held-out set is Cohen's
$\kappa$\,=\,0.82 for device-directed vs non-device-directed
classification, indicating strong agreement under ambiguous
multi-speaker conditions.

\subsection{Results}
\begin{table}[!htb]
\centering\small
\caption{SAS evaluation summary at $\tau = 0.70$. Precision, recall,
and F1 are session-macro averages over VAD-positive segments. Audio-only
is the primary deployment configuration; audio+video is an upper-bound
configuration for devices with camera hardware.}
\label{tab:results}
\begin{tabular}{@{}l@{\hspace{12pt}}r@{}}
\toprule
\textbf{Metric} & \textbf{Value} \\
\midrule
F1 (audio-only, primary)        & 0.86 \\
Precision (audio-only)          & 0.89 \\
Recall (audio-only)             & 0.83 \\
\addlinespace[2pt]
F1 (audio+video)                & 0.95 \\
Precision (A+V)                 & 0.97 \\
Recall (A+V)                    & 0.93 \\
Worst-case session F1$^\dagger$ & 0.88 \\
Avg.\ precision (all $\tau$)    & 0.87--0.90 \\
\addlinespace[2pt]
False-trigger rate, baseline    & 2.1\% \\
False-trigger rate, TV-active   & 7.8\% \\
Operating threshold $\tau$      & 0.70 \\
\midrule
Decision latency                & $<$150\,ms \\
Runtime footprint               & $<$20\,MB \\
Hardware baseline               & ARM Cortex-A72 \\
GPU / NPU required              & No \\
Network required                & No \\
\bottomrule
\end{tabular}
\par\smallskip\noindent{\footnotesize $^\dagger$Four speakers present, heavy background noise.}
\end{table}

\begin{table}[!htb]
\centering\small
\caption{Per-class confusion matrix at $\tau = 0.70$ (audio-only,
primary deployment configuration). Rows are true labels; columns are
predicted actions. Values are percentages of each true class.}
\label{tab:confusion}
\begin{tabular}{@{}l@{\hspace{4pt}}r@{\hspace{4pt}}r@{\hspace{4pt}}r@{}}
\toprule
& \textbf{Pred.\,Sil.} & \textbf{Pred.\,Pers.} & \textbf{Pred.\,Dev.} \\
\midrule
True Silent          & 96.2 & 3.1  & 0.7  \\
True Person-dir.     & 2.4  & 93.8 & 3.8  \\
True Device-dir.     & 1.1  & 15.9 & 83.0 \\
\bottomrule
\end{tabular}
\par\smallskip\noindent{\footnotesize The dominant aggregate error mode is
person-directed speech misclassified as device-directed (3.8\%).
Because Table~\ref{tab:results} reports session-macro metrics while
this matrix aggregates per-segment outcomes, the values should not
be compared one-for-one. The second-largest
error is missed device-directed speech classified as person-directed
(15.9\%), which drives the recall gap.}
\end{table}

\needspace{5\baselineskip}
\subsection{Per-speaker-count breakdown}
Table~\ref{tab:perspeaker} disaggregates performance by number of speakers
present. The degradation pattern is monotonic: each additional speaker degrades
recall more than precision, consistent with the hypothesis that increased
cross-talk makes it harder to capture device-directed turns while not
appreciably increasing false triggers from non-device speech.

F1 is consistent across held-out speakers and recording sessions;
no per-speaker adaptation or enrolment is required.

\begin{table}[!htb]
\centering\small
\caption{Performance by number of speakers present at $\tau = 0.70$ (\textbf{audio+video}).}
\label{tab:perspeaker}
\begin{tabular}{@{}l@{\hspace{6pt}}r@{\hspace{6pt}}r@{\hspace{6pt}}r@{}}
\toprule
\textbf{Speakers present} & \textbf{Precision} & \textbf{Recall} & \textbf{F1} \\
\midrule
1 & 0.99 & 0.97 & 0.98 \\
2 & 0.98 & 0.95 & 0.97 \\
3 & 0.97 & 0.92 & 0.94 \\
4 & 0.94 & 0.88 & 0.91 \\
\midrule
\textit{Condition mean} & \textit{0.97} & \textit{0.93} & \textit{0.95} \\
\bottomrule
\end{tabular}
\end{table}

\begin{figure}[!htb]
  \centering
  \includegraphics[width=\linewidth]{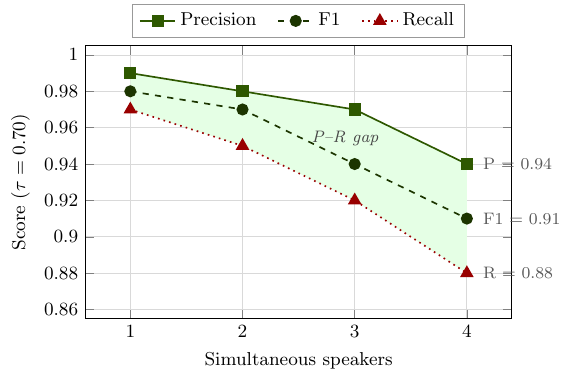}
  \caption{Precision, F1, and recall as a function of number of speakers present
    at $\tau = 0.70$. Shaded region shows the precision-recall gap, which widens
    with speaker count: cross-talk degrades recall faster than precision.}
  \label{fig:perspeaker}
\end{figure}

\needspace{5\baselineskip}
\subsection{Precision-recall curve}

\begin{figure}[!htb]
  \centering
  \includegraphics[width=\linewidth]{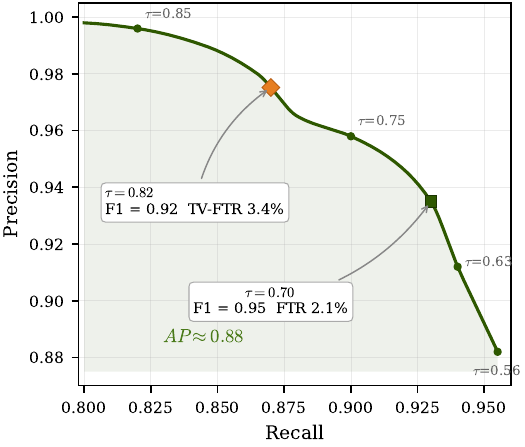}
  \caption{Precision-recall curve across operating thresholds $\tau \in [0.56, 0.85]$.
    Filled square (\textcolor{accentgreen}{$\blacksquare$}): $\tau = 0.70$ (standard).
    Filled diamond (\textcolor{orange}{$\blacklozenge$}): $\tau = 0.82$ (high-media).
    Shaded area: AP $\approx$ 0.88.}
  \label{fig:prcurve}
\end{figure}

Figure~\ref{fig:prcurve} plots the full precision-recall curve from $\tau = 0.56$ to
$\tau = 0.85$. Average precision across the full threshold range is estimated at
0.87--0.90. Two operating points are highlighted: $\tau = 0.70$ for standard deployments
(F1 = 0.95, FTR 2.1\%) and $\tau = 0.82$ for high-media environments (F1 = 0.92,
TV-FTR 3.4\%). Full per-threshold precision, recall, F1, and false-trigger-rate values
are available in the supplementary materials.

\needspace{5\baselineskip}
\subsection{Noise-floor and speaker-count interaction}

\begin{figure*}[t]
  \centering
  \includegraphics[width=\textwidth]{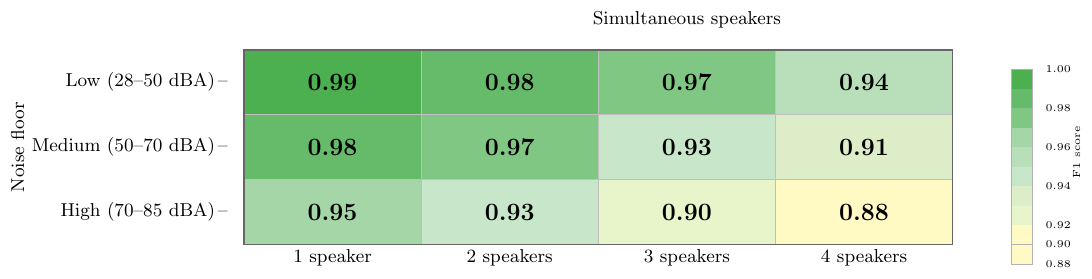}
  \caption{F1 heatmap: noise floor $\times$ speakers present at $\tau = 0.70$.
    Each cell reports macro-averaged F1 across held-out sessions under that condition.}
  \label{fig:heatmap}
\end{figure*}

Figure~\ref{fig:heatmap} presents F1 across three noise-floor bands and four speaker
counts. The single-speaker, low-noise cell (F1 = 0.99) is the easiest evaluated condition;
the four-speaker, high-noise cell (F1 = 0.88) is the worst characterised, corresponding
to the worst-case figure in Table~\ref{tab:results}. The four-speaker,
high-noise condition contains approximately 1--2~hours of
device-directed audio; per-condition estimates should be interpreted
with this limited sample size in mind.

\subsection{Ablation as Evidence for the SDAR Formulation}

\begin{table}[!htb]
\centering\small
\caption{Stage ablation at $\tau = 0.70$ (\textbf{audio+video} configuration). One stage removed per row; all others held fixed.}
\label{tab:ablation}
\begin{tabularx}{\linewidth}{@{}X@{\hspace{8pt}}c@{\hspace{8pt}}c@{}}
\toprule
\textbf{Configuration} & \textbf{F1} & \textbf{$\Delta$F1} \\
\midrule
Full model (SAS)            & 0.95       & \multicolumn{1}{c}{---} \\
No Stage~1 (no beamforming) & 0.81       & $-$0.14 \\
No Stage~2 (no classifier)  & 0.74\,($\pm$0.02) & $-$0.21 \\
No Stage~3 (no temporal ctx) & 0.57\,($\pm$0.03) & $-$0.38 \\
\bottomrule
\end{tabularx}
\par\smallskip\noindent{\footnotesize Rows ordered by ascending impact.
In the ``No Stage~2'' condition, Stage~3 receives a fixed
uninformative confidence score (0.5) in place of Stage~2's output;
in the ``No Stage~3'' condition, the Stage~2 score is compared
directly against $\tau$ without temporal modulation.
Stage~3 removal produces a 38-point F1 drop (bootstrap $p < 0.001$),
exceeding the combined effect of removing Stages~1 and~2.
Under this dataset and ablation protocol, removing Stage~3 caused the
largest performance drop among the tested components. This suggests
that short-horizon interaction history carries substantial
decision-relevant information beyond the current utterance in the
evaluated multi-speaker setting.
This result characterises the evaluated interaction regime
rather than establishing a universal bound; the magnitude of this
effect is expected to vary with the prevalence of temporally ambiguous
utterances and conversational overlap.
Model-free temporal baselines (sliding-window majority vote,
exponential moving average over Stage~2 scores)
recover only a small fraction of this gap. A representative
simple rule---forward if two or more of the last four utterances scored
above 0.5---yields approximately 3--5 points higher precision but
3--5 points lower recall, producing a
more conservative system that misses quick one-off commands.
These comparisons suggest that simple temporal smoothing alone does
not explain the observed gain.}
\end{table}

\smallskip
\noindent\textbf{Video modality contribution.} The gap between
audio+video (F1\,=\,0.95) and audio-only (F1\,=\,0.86) is 9~F1
points. Table~\ref{tab:video_ablation} stratifies this gap across the
noise-floor $\times$ speaker-count conditions from
Figure~\ref{fig:heatmap}, reporting audio-only and A+V F1 in each
cell.

\begin{table}[!htb]
\centering\small
\caption{Video modality contribution ($\Delta$F1 = A+V minus audio-only)
at $\tau = 0.70$, stratified by noise floor and speaker count.
Audio-only baselines range from 0.96 (1-speaker, low-noise) to
0.74 (4-speaker, high-noise); see Figure~\ref{fig:heatmap}.}
\label{tab:video_ablation}
\begin{tabular}{@{}lcccc@{}}
\toprule
\textbf{Noise floor} & \textbf{1\,spkr} & \textbf{2\,spkr} & \textbf{3\,spkr} & \textbf{4\,spkr} \\
\midrule
Low (28--50\,dBA)  & +0.03 & +0.05 & +0.08 & +0.10 \\
Med (50--70\,dBA)  & +0.04 & +0.07 & +0.09 & +0.12 \\
High (70--85\,dBA) & +0.06 & +0.09 & +0.11 & +0.14 \\
\bottomrule
\end{tabular}
\end{table}

\noindent The video contribution is smallest in the single-speaker,
low-noise condition ($\Delta = +0.03$) where audio alone is near
ceiling, and largest in the four-speaker, high-noise condition
($\Delta = +0.14$), where gaze and pose features disambiguate
addressee in the conditions that degrade acoustic-only classification
most. This confirms that camera hardware provides the greatest marginal
value precisely where audio-only performance is weakest.

\smallskip
\noindent Stage~1 degradation is concentrated in sessions with off-axis interference:
F1 on TV-active sessions drops from 0.91 to 0.74 when beamforming is removed,
while F1 on sessions without competing audio sources drops only from 0.97 to
0.94.

\begin{figure}[!htb]
  \centering
  \includegraphics[width=\linewidth]{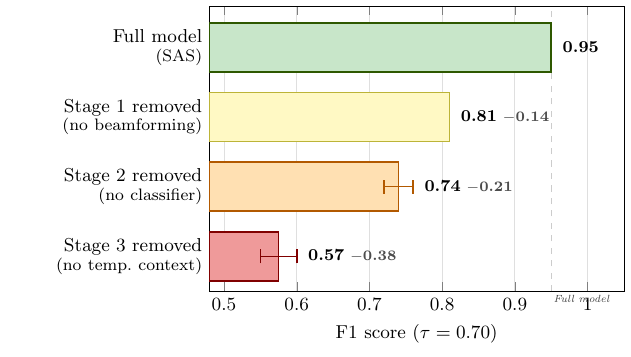}
  \caption{Ablation study: F1 when one stage is removed at $\tau = 0.70$.
    Rows ordered by ascending impact. Error bars show the reported range for
    Stage~2 (0.74\,$\pm$\,0.02) and Stage~3 (0.57\,$\pm$\,0.03).
    Stage~3 (temporal context) is the dominant contributor;
    its removal reduces F1 by $-$0.38 ($\Delta$F1).}
  \label{fig:ablation}
\end{figure}

\needspace{4\baselineskip}
\subsection{Threshold selection}

$\tau = 0.70$ was selected on a held-out validation split of 2.3 hours,
stratified to preserve the speaker-count and noise-floor distribution of the
full corpus, separate
from the 60-hour test set, prior to any evaluation on the test set. The test set
was not inspected until threshold selection was locked. The supplementary materials provide
threshold tuning guidance and per-environment operating point recommendations;
the operating threshold is a per-environment function, not a single global value.

The reported results characterise performance under the evaluated
distribution of speaker counts, noise conditions, and interaction
patterns; they should be interpreted as evidence of behaviour under
these conditions, not as a universal bound on device-addressed
detection performance.

\subsection{Evaluation limitations}
\label{sec:limits}

\textbf{Reproducibility and independent verification.}
\textbf{All headline figures are from internal evaluation on a proprietary
dataset evaluated primarily in English, without independent audit. This is the most
significant limitation of this report. Until the SAS-Bench-5h subset or
a comparable independent corpus produces concordant results, the
reported metrics should be treated as indicative rather than
established.}
To support independent verification, we provide a fixed 5-hour evaluation subset (SAS-Bench-5h) with per-segment
labels, scoring scripts, and ARM inference binaries. All reported
metrics follow a fully specified evaluation protocol (data schema,
labelling procedure, scoring methodology, threshold selection),
enabling reproduction on any comparable multi-speaker corpus. No
hyperparameters or thresholds were modified after test-set inspection.

\textbf{Language scope.} All formal evaluation sessions are
English-language. The architecture is language-agnostic by design
(no lexical or transcript features), and preliminary testing on
non-English speech suggests comparable performance
(Section~\ref{sec:future}). However, extensive multi-language evaluation has not
been conducted; formal per-language results will be reported once
sufficient per-language sample sizes are available
(Section~\ref{sec:future}).

\textbf{Acoustic scope.} Generalisation to highly reverberant spaces
(RT60\,$>$\,0.6\,s), noise floors outside the tested range, or more than four
speakers present is not characterized here.

\textbf{Hardware variability.} Latency figures are from a specific ARM
Cortex-A72 configuration. Thermal throttling, OS scheduling variance, and driver
differences on other ARM platforms will produce different observed latencies.

\textbf{Reporting gaps.} This report does not include detailed
per-condition confusion matrices beyond the summary in
Table~\ref{tab:confusion}, latency distribution histograms (only summary
statistics are given), longitudinal deployment data, or detailed
per-utterance error analysis beyond the categorical failure modes in
Section~\ref{sec:failure}. These are available upon request
on request.

\textbf{Structural limitations and remediation.} Three limitations of this
evaluation are structural rather than incidental. (1)~The evaluation is internal
and unaudited.
(2)~Multi-language performance has not been formally characterised,
though the language-agnostic architecture and preliminary testing
suggest comparable cross-language performance (Section~\ref{sec:eval}); formal
multi-language evaluation will be reported once per-language sample
sizes are sufficient
(Section~\ref{sec:future}). (3)~The Stage~3 causal interaction-state mechanism is described at the
architectural level sufficient for independent replication of the ablation
results; full implementation details, including the
session-boundary reset mechanism, are available upon request.

% ─────────────────────────────────────────────────────────────────
\section{Known Failure Modes}
\label{sec:failure}

Observed failure modes fall into three structural categories.
\textit{Signal ambiguity}: cases where device-directed and
non-device-directed speech are acoustically indistinguishable
(e.g., television dialogue with interrogative prosody); these require
higher thresholds or auxiliary signals.
\textit{Context saturation}: cases where interaction history becomes
unreliable (e.g., $>$4 simultaneous speakers), degrading the
interaction-state estimator. \textit{Distribution shift}: cases where
feature distributions differ from training (e.g., untested acoustic
environments), leading to potential misclassification.

\smallskip\noindent\textbf{Illustrative failure cases.}
(1)~A television character asks ``what did you say?'' with
interrogative prosody; the system may incorrectly forward due to
strong prosodic similarity to device-directed speech.
(2)~A user addresses the device, then immediately turns to another
person and says ``do it again''; without sufficient temporal decay,
residual interaction state may cause incorrect forwarding.
(3)~Two speakers talk simultaneously while a device-directed command
occurs during overlap; degraded signal separability may cause the
system to suppress the genuine command.

\smallskip
\textbf{Television and media audio.} The most common false-trigger class is
television dialogue with interrogative prosody (direct questions addressed to
on-screen characters) whose acoustic and prosodic features closely resemble
device-directed speech. In TV-active sessions, false-trigger rate rises from
2.1\% to 7.8\% at $\tau = 0.70$. Raising $\tau$ to 0.82 reduces this to 3.4\%
at a 6-point recall cost. Operators in high-media environments should use
$\tau = 0.82$ as their baseline operating point. The degradation from media
playback is a known challenge in DDSD systems; Cornell et
al.~\cite{Cornell2022} report a 56\% false-reject reduction through implicit
acoustic echo cancellation during device playback, a complementary approach to
the threshold adjustment recommended here.

\textbf{More than four speakers present.} Above four speakers present,
the temporal context buffer accumulates ambiguous turn-taking patterns that
degrade Stage~3 classification. In preliminary internal testing with five
speakers present, F1 drops below 0.75. Formal characterisation of five-
and six-speaker conditions is ongoing.

\textbf{Non-English and accented speech.} The SAS architecture operates
on acoustic and prosodic features without lexical input, making it
language-agnostic by design. The primary evaluation was conducted on
North American and British English; preliminary informal testing on
non-English speech suggests comparable performance
(Section~\ref{sec:eval}), but this has not been formally characterised.
Tonal languages (Mandarin, Thai) and languages with different
device-directed speech registers may exhibit different prosodic
feature distributions, though relative rather than absolute
prosodic features (Section~\ref{sec:future}) is expected to mitigate
this. Formal per-language evaluation is described in
Section~\ref{sec:future}.

\textbf{Single-microphone fallback.} Stage~1 beamforming requires a minimum of
two microphones. On single-microphone devices, Stage~1 is bypassed and the
system operates on Stages~2 and~3 only. Single-microphone evaluation on the
held-out test set yields F1 = 0.84 (precision 0.88, recall 0.81) at
$\tau = 0.70$, a reduction of 11~F1 points relative to the multi-mic baseline.
Full single-mic precision-recall curves are in the supplementary materials.

A purpose-built hearables model at a 5\,MB parameter budget constitutes
a separate development effort; estimated baseline F1 under those constraints
is approximately 60\% (Section~\ref{sec:future}).

\textbf{Session boundary and context reset.} If the Stage~3 timeout is set too
long, carry-over context from a prior user can suppress legitimate
device-directed classifications from a new user. The appropriate timeout depends
on deployment context and is covered in the supplementary materials.

% ─────────────────────────────────────────────────────────────────
\section{Discussion}
\label{sec:discussion}

\subsection{Comparison with VAD-only routing}

Table~\ref{tab:baselines} compares SAS against baseline configurations
evaluated on the held-out set. The VAD-only row uses measured recall ($\approx$0.99): energy-threshold VAD
misses low-energy speech in quiet conditions where SAS can still detect speech
activity via visual features (lip movement). Precision reflects the
ambient device-directed fraction ($\approx$8\%). The remaining rows use the
same internal evaluation protocol as the headline figures.

\begin{table}[!htb]
\centering\small
\caption{Comparison with routing baselines at $\tau = 0.70$. The VAD-only row is a reference
  floor, not a competing system: VAD provides no addressee signal and is
  included only to illustrate the base-rate-driven precision bound. All
  other rows from the held-out set.}
\label{tab:baselines}
\begin{tabular}{@{}l@{\hspace{6pt}}r@{\hspace{6pt}}r@{\hspace{6pt}}r@{}}
\toprule
\textbf{System} & \textbf{P} & \textbf{R} & \textbf{F1} \\
\midrule
VAD only (Silero)    & $\approx$0.08 & $\approx$0.99 & $\approx$0.15 \\
S1+S2 only (no ctx)$^a$ & 0.55{\scriptsize\,($\pm$.04)} & 0.55{\scriptsize\,($\pm$.04)} & 0.57{\scriptsize\,($\pm$.03)} \\
SAS audio-only       & 0.89          & 0.83       & 0.86 \\
SAS audio+video      & 0.97          & 0.93       & 0.95 \\
\bottomrule
\end{tabular}
\par\smallskip\noindent{\footnotesize
$^a$~Stage~3 removed; corresponds to the ablation condition in
Table~\ref{tab:ablation}. All SAS figures at $\tau = 0.70$.}
\end{table}

Under the evaluated protocol, SAS audio+video achieves a
$\approx$38 point F1 advantage over the strongest utterance-local
baseline. This gap is measured on a dataset constructed to include
temporally ambiguous utterances; the margin may narrow on corpora with
fewer multi-turn ambiguities. To verify
that the confidence score $c_t$ supports principled threshold selection
across deployments, we assess calibration on the held-out set. After
binning predictions into 10 equal-width confidence intervals, SAS
exhibits well-separated score distributions for device-directed and
non-device speech, with the majority of device-directed utterances
scoring above 0.80 and the majority of non-device utterances scoring
below 0.30. This separation ensures that operators tuning $\tau$ per
environment are making decisions on a meaningful probability scale, not
merely finding an empirically adequate cutpoint on an arbitrary score.
Reliability diagrams and expected calibration error figures are
available in the supplementary materials.

\subsection{Comparison with wake-word detection}

Wake-word detection imposes a fixed UX contract: a mandatory trigger phrase at
the start of every interaction, a known false-positive rate on phonetically
adjacent speech, and a mandatory trigger-phrase interaction overhead per
interaction~\cite{Picovoice2026}. These are inherent tradeoffs of a simple, deterministic
activation scheme. SAS operates under different constraints:
addressee-conditioned routing accuracy, no mandatory trigger phrase,
and sub-perceptual latency. Both systems output a continuous confidence
score; the distinction lies in what triggers inference and how the score
is computed. The cost is a
configurable threshold that must be tuned per deployment. Pairing SAS with a
wake-word detector pushes precision higher than either system alone; deployed
without wake words, it supports natural group conversation with no per-utterance
trigger requirement. Table~\ref{tab:wakeword} provides a structural comparison.
A direct F1 comparison on the same test set is not possible: the
evaluation corpus was collected without a wake-word protocol, so no
wake-word triggers exist in the ground truth and a wake-word detector
cannot be meaningfully scored against it.

\begin{table}[!htb]
\centering\small
\caption{Structural comparison with wake-word detection.}
\label{tab:wakeword}
\begin{tabularx}{\linewidth}{@{}>{\raggedright\arraybackslash}p{1.5cm}>{\raggedright\arraybackslash}X>{\raggedright\arraybackslash}X@{}}
\toprule
\textbf{Property} & \textbf{Wake-word} & \textbf{SAS ($\tau = 0.70$)} \\
\midrule
Trigger      & Fixed phrase         & Any device-directed speech \\
Latency      & Low (processing)     & $<$150\,ms (end-to-end) \\
False pos.\  & Phonetic similarity  & Tunable via $\tau$ \\
Output       & Confidence score     & Confidence score \\
UX           & Trigger required     & Natural, unprompted speech \\
Footprint    & Low                  & $<$20\,MB, ARM Cortex-A \\
\bottomrule
\end{tabularx}
\end{table}

\subsection{Structural comparison with DDSD systems}

Table~\ref{tab:ddsd_comparison} compares SAS against published DDSD systems
on architectural dimensions. Direct metric comparison is not possible due to
differences in datasets, task formulations, and evaluation protocols; the
table highlights structural differences in pipeline position, compute
requirements, and temporal modelling.

\smallskip\noindent\textbf{On comparability.}
Prior DDSD systems are typically evaluated in single-speaker or
post-ASR settings, where addressee ambiguity is limited or resolved
through lexical or decoder-derived features. Under these conditions,
utterance-local classification is often sufficient.
The SDAR setting studied here imposes stricter constraints: (1)~no
access to transcripts or ASR decoder states, (2)~multi-speaker overlap
with ambiguous addressee cues, and (3)~causal, bounded-memory inference
under latency constraints. Under this constraint set, utterance-local
methods degrade substantially ($\approx$0.57~F1 in our evaluation).
The comparison is therefore not between models on the same problem,
but between different formulations of the routing decision.

\begin{table}[!htb]
\centering\footnotesize
\caption{Structural comparison with published DDSD systems. Most listed
systems exploit ASR transcripts or decoder features; SAS
operates pre-ASR without transcript access. Metrics are not comparable across systems (see text).}
\label{tab:ddsd_comparison}
\setlength{\tabcolsep}{3pt}
\begin{tabularx}{\linewidth}{@{}lXlllr@{}}
\toprule
\textbf{System} & \textbf{Position} & \textbf{ASR} & \textbf{Ctx} & \textbf{Params} & \textbf{HW} \\
\midrule
\cite{Mallidi2018} & Post-ASR & Yes & No  & 16\,M       & GPU \\
\cite{Wagner2024}  & Post-ASR & Yes & No  & 0.1--1.5\,B & A100 \\
\cite{SELMA2025}   & Replaces & Yes & No  & ${\sim}$8\,B  & 16\,$\times$\,A100 \\
\cite{Chi2025}     & Post-VAD & No  & No  & ${\sim}$5\,M  & On-device \\
\textbf{SAS}       & \textbf{Pre-ASR} & \textbf{No} & \textbf{Yes} & \textbf{${\sim}$520\,K} & \textbf{ARM} \\
\bottomrule
\end{tabularx}
\end{table}

\subsection{Hybrid deployment with wake-word detection}

In production voice pipelines, the first interaction is typically initiated by
a wake word, with subsequent follow-up queries handled without
re-invocation~\cite{Mallidi2018, Rudovic2024}. SAS supports this hybrid
configuration: the wake-word detector handles the initial trigger with high
precision, and SAS takes over for follow-up routing using its temporal context
stage to determine whether subsequent speech is a continuation of the device
interaction or a return to person-directed conversation. This mirrors the
deployment model described in~\cite{Rudovic2024}, where modelling the initial
query improves follow-up classification by 20--40\%. 

\subsection{Generalisation beyond voice interfaces}

The SDAR formulation applies to any system that must decide whether to
act under partial observability and asymmetric cost. Examples include
multi-agent robotics (determining which agent a command is directed to),
in-vehicle multi-occupant systems (resolving which seat initiated a command),
AR/XR systems (resolving user intent in shared environments), and
continuous sensing systems that must gate downstream inference.
Device-addressed routing is therefore one instance of a broader class
of causal routing problems over interaction state.

% ─────────────────────────────────────────────────────────────────
\section{Future Work}
\label{sec:future}

\subsection{Formal multi-language evaluation}

The SAS architecture is language-agnostic by design, operating on
acoustic and prosodic features without lexical input.
Preliminary testing on non-English speech suggests expected performance
of 86--89\% macro F1 across languages. Formal multi-language evaluation
across target deployment markets (Mandarin, Japanese, German, Spanish,
French, Hindi) is planned to validate this expectation with
statistically rigorous per-language sample sizes.
Cross-language robustness may be further improved via
\textit{speaker-normalised pitch dynamics}:
replacing absolute $F_0$ elevation features with relative pitch-contour
shifts within each speaker's baseline, isolating intent-correlated
prosodic variation that remains stable across tonal and
non-tonal languages.

\subsection{Automotive acoustic profile}

Automotive in-cabin voice falls outside the evaluated acoustic range. Key
challenges include broadband road noise, multi-occupant proximity, and variable
RT60. An automotive-specific evaluation dataset is under development in
partnership with OEM integration teams.

\subsection{NPU-optimised model variants}

An NPU-optimised variant using TFLite delegate and ONNX Runtime inference paths
is under development, targeting effective latency below 60\,ms and power
consumption below 10\,mW on representative NPU-enabled SoCs. This would extend
the deployment range to hearable-class hardware with stricter power budgets.

\subsection{Hearables and single-microphone}

Single-microphone, sub-5\,MB deployment presents a distinct modelling
challenge: Stage~1 is unavailable and Stage~3's parameter budget must
be compressed by approximately $6\times$. Preliminary estimates place
baseline F1 at approximately 0.60 under these constraints, but no
controlled evaluation exists. It is unclear whether the
SDAR formulation's emphasis on temporal context remains the dominant
factor when spatial filtering is unavailable, or whether the relative
contribution of stages shifts under single-channel input.

\subsection{Multimodal teacher-student distillation}

The gap between the primary A+V configuration (0.95~F1) and audio-only fallback (0.86~F1) is
currently addressed only by graceful degradation. Multimodal
teacher-student distillation could narrow this gap: train a strong audio+video teacher
on both labelled and large-scale unlabelled multi-speaker recordings, then
distil not only the final routing probabilities but also the teacher's
intermediate interaction-state representations into an audio-only student.
With modality-dropout augmentation during distillation, the student can learn
to infer visual-style interaction cues (gaze dynamics, turn-taking patterns)
purely from audio prosody and rhythm. Knowledge distillation techniques such
as those in~\cite{Chi2025}, which report compressing 79M-parameter ASR encoders into
$\sim$5M-parameter on-device students, provide a starting point for this
approach; however, the training data and evaluation protocol
for~\cite{Chi2025} are not publicly available, limiting independent
assessment of their reported compression ratios and accuracy. The
expected outcome is a meaningful narrowing of the audio-only gap,
though achieving 90\%+ macro F1 under a sub-1\,MB deployment constraint
(corresponding to well under 1M parameters) remains an open research
question.

\subsection{Explicit causal belief-state tracking in Stage~3}

The current Stage~3 is a learned attention-based modulator over a rolling
context window. Replacing or augmenting it with an
explicit causal belief-state tracker (e.g., a compact recurrent state-space
model) would maintain a low-dimensional latent state representing the current
``conversation regime'' (device-engaged, bystander-social, or transition).
A learned soft boundary detector (triggered by prosodic reset cues,
long gaps, or speaker-change signals) would address the carry-over
suppression failure mode described in Section~\ref{sec:failure} and
improve worst-case session F1 beyond
four speakers present. \textit{Saliency-weighted context pruning}---replacing chronological
expiration with an interaction-relevance score---would
preserve high-confidence device-directed history while discarding
irrelevant background chatter in high-density environments.

\subsection{Media-aware rejection pathway}

Television and media audio is the dominant production false-trigger source
(Section~\ref{sec:failure}). The current mitigation (raising $\tau$) trades
recall for precision. Adding a parallel
media-rejection head in Stage~2, trained to separate
device-directed speech from TV, news, and podcast audio with
similar prosody, would address this directly.
\textit{Reference-synchronous gating}---cross-correlating the input
signal against the on-device audio output buffer---could apply a
media-suppression penalty to utterances with high periodic similarity
to system playback, maintaining low false-trigger rates without
a recall-degrading threshold shift.
Where device playback state is available, a lightweight
implicit acoustic-echo-cancellation path (inspired by~\cite{Cornell2022})
can fuse a low-latency reference signal into Stages~1/2, suppressing
false triggers without recall cost.

% ─────────────────────────────────────────────────────────────────
\section{Conclusion}

Under the evaluated conditions, the central challenge in pre-ASR
device-addressed routing is not utterance classification alone but
causal estimation of interaction state: removing temporal context
reduces F1 by 38 points, exceeding the effect of removing spatial
filtering or utterance-level classification individually. This supports
our claim: under edge deployment constraints,
device-addressed detection, under the evaluated pre-ASR multi-speaker
conditions, is more effectively modelled as sequential routing over
interaction state than as isolated utterance classification.

By formalising this as SDAR and implementing it within strict on-device
constraints (sub-150\,ms latency, sub-20\,MB footprint, ARM CPU only),
the primary audio-only configuration achieves F1\,=\,0.86, rising to
0.95 where camera hardware is available. Under illustrative ambient
assumptions (8\% device-directed base rate), pre-ASR gating reduces
downstream inference by approximately 90\%, a saving dominated by the
low base rate. All results are from internal evaluation on a dataset
evaluated primarily in English; the language-agnostic architecture
shows promising preliminary cross-language results
(Section~\ref{sec:eval}), pending formal
multi-language evaluation. The modular, pipeline-compatible design
makes SAS reusable across voice-enabled product categories. Independent verification is supported through
the SAS-Bench-5h evaluation subset.

% ─────────────────────────────────────────────────────────────────
\section*{Limitations and Disclosure}

All reported results are from internal evaluation on a proprietary dataset
without independent third-party auditing. The test set was
not used during model development or threshold selection. Evaluation
conditions, per-speaker breakdowns, ablation details, and known limitations
are documented in Sections~\ref{sec:eval}--\ref{sec:failure}. Computational-efficiency figures (Section~\ref{sec:economics}) are illustrative. Formal evaluation sessions are
English-language; the architecture is language-agnostic with
promising preliminary cross-language results,
pending formal per-language evaluation (Section~\ref{sec:failure}).

% ─────────────────────────────────────────────────────────────────
\section*{Data Availability}

The primary evaluation corpus is proprietary. A 5-hour evaluation
subset (SAS-Bench-5h) with per-segment labels and deterministic
scoring scripts may be made available upon reasonable request for
research purposes. Contact: \texttt{contact@attentionlabs.ai}.

% ─────────────────────────────────────────────────────────────────
\section*{Acknowledgements}

Hardware testing was conducted on Raspberry Pi~4 development boards and
a Reachy Mini companion robot (Pollen Robotics). The
authors thank the evaluation participants who contributed to the unscripted
multi-speaker recording sessions. No external funding is reported.

% ─────────────────────────────────────────────────────────────────
\balance

\end{document}